\documentclass[11pt]{article}

\usepackage{cite}
\usepackage{color}
\usepackage{amsfonts}
\usepackage{amsmath}
\usepackage{epsfig}
\usepackage{latexsym}
\usepackage{fancybox}
\usepackage{graphicx}
\usepackage{subfigure}
\usepackage{amssymb}

\numberwithin{equation}{section}
\allowdisplaybreaks

\def\spa#1{\phantom{\fbox{\rule[-#1cm]{0cm}{0cm}}}}

\newcommand{\rf}[1]{(\ref{#1})}
\newcommand{\beq}{\begin{equation}}

\newcommand{\eeq}{\end{equation}}
\newcommand{\bea}{\begin{eqnarray}}
\newcommand{\eea}{\end{eqnarray}}

\newcommand{\vp}{\varepsilon}

\def\[#1\]{\begin{align}#1\end{align}}

\begin{document}

\hfuzz=100pt
\title{{\Large \bf{
Towards elucidation of zero-temperature criticality of\\
Ising model on 2d DT 
}}}
\author{Jan Ambj\o rn$^{a,b}$\footnote{ambjorn@nbi.dk}, Yuki Sato$^{c,d}$\footnote{ysato@th.phys.nagoya-u.ac.jp } and Tomo Tanaka$^{e}$\footnote{tomo@gravity.phys.waseda.ac.jp}
  \spa{0.5} \\
\\
$^a${\small{\it The Niels Bohr Institute, Copenhagen University}}
\\ {\small{\it Blegdamsvej 17, DK-2100 Copenhagen, Denmark}}\\    
\\
$^b${\small{\it IMPP, Radboud University}}
\\ {\small{\it Heyendaalseweg 135, 6525 AJ, Nijmegen, The Netherlands}}\\    
\\
$^c${\small{\it Institute for Advanced Research, Nagoya University}}
\\ {\small{\it Chikusaku, Nagoya 464-8602, Japan}}\\  
\\
$^d${\small{\it Department of Physics, Nagoya University}}
\\ {\small{\it Chikusaku, Nagoya 464-8602, Japan}}\\  
\\
$^{e}${\small{\it Department of Physics, Waseda University}}
\\ {\small{\it Okubo, Tokyo, 169-8555, Japan}}
\spa{0.3} 
}
\date{}

\maketitle
\centerline{}

\begin{abstract} 
We study the zero-temperature criticality of the Ising model on two-dimensional dynamical triangulations 
to contemplate its physics. 
As it turns out, an inhomogeneous nature of the system yields an interesting phase diagram and the physics at the zero temperature 
is quite sensitive about how we cool down the system.   
We show the existence of a continuous parameter that characterizes the way we approach the zero-temperature critical point 
and it may enter in a critical exponent.       
\end{abstract}

\renewcommand{\thefootnote}{\arabic{footnote}}
\setcounter{footnote}{0}

\newpage

\section{Introduction}
\label{sec:introduction}

The Ising model on $2$d dynamical triangulations (DT) is a statistical system including quantum effects of gravity, 
which was studied first by Boulatov and Kazakov \cite{Kazakov:1986hu, Boulatov:1986sb}\footnote{
DT was first introduced as a regularization of $2$d quantum gravity \cite{Ambjorn:1985az,Ambjorn:1985dn,David:1984tx,Billoire:1985ur,Kazakov:1985ea,Boulatov:1986jd}.  
For a review see  \cite{Ambjorn:1997di}.    
}. We call this model the Boulatov-Kazakov Ising model. 
Through the use of matrix models, the Boulatov-Kazakov Ising model was solved exactly 
and all the critical exponents were obtained analytically \cite{Boulatov:1986sb}. 
Those are different from Onsager's critical exponents of the Ising model on a flat $2$d regular lattice \cite{Onsager:1943jn}.  
Due to quantum gravitational effects, the phase transition is changed from a second order transition  to a
 third order transition at a finite critical temperature, 
and the continuum theory defined at the critical temperature turns out to be  Liouville gravity  coupled to a conformal 
field theory  with  central charge $c = 1/2$. 
The scaling dimensions of primary operators in the $c=1/2$ conformal field theory change to the Knizhnik-Polyakov-Zamolodchikov (KPZ) values \cite{Knizhnik:1988ak}, which are caused  by  gravitational dressing.
At the same time the long range fluctuations of spins  interacting with geometry changes 
the critical properties of 2d quantum gravity itself at the critical temperature.     

Causal dynamical triangulations (CDT) were introduced as a different class of triangulations \cite{al,ajl,ajl1}, 
mainly in an attempt  to cure some of the problems encountered in the DT formalism in more than 2 dimensions
(see  \cite{Ambjorn:2012jv,lollreview} for reviews).  However, one can also study 2d CDT, and in particular one can
couple the Ising spins to the model, in the same spirit as described above. This spin model has not been solved
analytically, but it can be studied by high temperature expansions and Monte Carlo simulations and the 
results are clear: the critical exponents of the Ising model coupled to CDT are identical to the Onsager exponents
\cite{aal,aal1}.  The 2d  CDT model allows much less geometrical fluctuations than the 2d DT model and the allowed 
fluctuations are not strong enough to change the Onsager exponents of the spin system\footnote{While the 
critical spin exponents  remain the ones of the flat space, the back reaction of spins on the geometry 
in the case of more than two  Ising models coupled the triangulations is quite strong. Thus we have a kind of
$c=1$ barrier even for CDT. This is confirmed by studying massless scalar  fields coupled to CDT \cite{gaussian}} . 
In particular
it has been shown that the ability to create baby universes are important for the change of the Onsager 
exponents, and this is explicitly forbidden in 2d CDT \cite{adjspin}.

A generalized 2d CDT model (GCDT) which allowed the creation of a finite number baby universes 
was introduced first as a continuum theory \cite{Ambjorn:2007jm, Ambjorn:2008ta} and later defined at the discrete level \cite{Ambjorn:2008gk, Ambjorn:2013csx}. 
As clarified in \cite{Ambjorn:2013csx}, the difference between the lattice structures of GCDT and DT is the presence of 
a weight $\theta$ in GCDT that controls the number of baby universes, 
i.e. the number of local maxima of the distance labeling from a vertex picked up by hand for the sake of convenience.  
In GCDT at the discrete level, one can take the continuum limit of DT for a fixed $\theta>0$ where the number of baby universes diverges,  
while tuning $\theta \to 0$ one can reach the continuum limit of GCDT characterized by a finite number of baby universes, 
which includes the continuum CDT if there exits a unique local maximum, i.e. a global maximum, of the distance labeling.

While the graphs used to define GCDT at a discretized level can be considered as a relative small extension of the 
graphs used to define CDT, there exists a bijection between the ensemble of graphs defining GCDT and a set of graphs 
characterized by having a  finite number of faces.  The bijection is such 
that the number of baby universes in GCDT is precisely the number of faces in this other set of graphs\cite{Ambjorn:2013csx}. 
These latter graphs thus  consist of (infinitely) many tree-subgraphs and a finite number of faces in a specific continuum limit corresponding to the continuum limit of GCDT. 
Note here that when it comes to aspects of pure $2$d quantum gravity (i.e.\ gravity without matter fields), 
even though the lattice structures of these latter graphs and those of GCDT are quite different, 
since one just basically counts the entropy of the graphs, these two classes of graphs lead to the same theory due to the bijection.
However, these two classes of graphs will not necessarily  lead to the same theories when matter is coupled to the graphs.
This is illustrated in the case of CDT. If one couples Ising spin to the (rather regular) graphs originally used to define CDT,
the Ising model will behave more or less like Ising spins on a regular lattice and in particular there is a phase transition 
with Onsager critical exponents, as mentioned above. However, via the bijection these CDT graphs are mapped to graphs
with just one face, i.e.\ they are basically tree-graphs. It is known that Ising spins on tree graphs cannot be critical.

One advantage of studying GCDT through the graphs consisting of tree-subgraphs and a finite number of faces is
that there exists a one-matrix model with a cubic interaction and a tadpole term which allows us 
to introduce the parameter $\theta$ mentioned above and
which  can interpolate between DT and GCDT (realized on the set of graphs with a finite number of faces ) \cite{Ambjorn:2008jf,Ambjorn:2008gk}. It also allowed for 
an intuitive understanding of this interpolation in terms of an inhomogeneous lattice structure, 
as well as the possibility of new scaling limits using this inhomogeneous lattice structure \cite{Ambjorn:2014bga}.

It is possible to couple Ising spins to GCDT (realized via the set of graphs with a finite number of faces) 
in the spirit of Kazakov and Boulatov, using a two-matrix model. 
It was first done in \cite{Fuji:2011ce}\footnote{
This model is inequivalent to the Ising model on the original GCDT including a finite number of baby universes,  
since the information on Ising-spin configurations is not preserved through the bijection.  
}. 
The corresponding two-matrix model was explicitly solved and shown to be related to the Boulatov-Kazakov model in  
\cite{Sato:2017ccb}. In the one-matrix model \cite{Ambjorn:2014bga} it was shown, as mentioned above, 
how  a scaling parameter 
$\theta$ allowed one to reach the GCDT regime from the DT regime in the limit $\theta \to 0$. These considerations
were extended to the Ising model in the two-matrix model of \cite{Sato:2017ccb}. For a fixed  scaling parameter 
$\theta > 0$ the model can be mapped onto the Boulatov-Kazakov Ising model. The critical temperature of the Ising
spins of the model is a function of $\theta$ and for $\theta \to 0$ this critical temperature also goes to zero (but 
for any $\theta > 0$ it is mapped to the critical temperature of the Boulatov-Kazakov Ising model, which is of course 
independent of $\theta$). Thus the limit $\theta \to 0$ is interesting, since the geometry of the triangulations 
might change from DT to GCDT. According to \cite{Ambjorn:2014bga} there are even several ways to 
take the $\theta \to 0$ limits, leading 
to different ensembles of triangulations with different fractal properties, and consequently these different limits might also 
lead to different critical behavior of the Ising spins. This is the topic we want to study in this article.

 This paper is organized as follows. 
 In section \ref{sec:BoulatovKazakov}, we review the Boulatov-Kazakov Ising model in a self-contained manner. 
 Section \ref{sec:asetupforcooling} is devoted to an introduction of the model \cite{Sato:2017ccb} and in particular to
 explain useful tools we use in due course. 
 We then study the critical behaviors of the system in section \ref{sec:criticality}, focusing especially on the zero temperature. Section \ref{sec:discussion} contains summary and discussion.

\section{Boulatov-Kazakov Ising model}
\label{sec:BoulatovKazakov}
The Ising model on $2$d dynamical triangulations (DT) was first introduced in the seminal paper by Kazakov \cite{Kazakov:1986hu}, 
and with an external magnetic field added to  the system, all the critical exponents of the model 
could be  calculated analytically \cite{Boulatov:1986sb}. 
In this section we give a short review of this model  (the Boulatov-Kazakov Ising model) without 
 an external magnetic field, using triangulations with the Ising spins placed in the center of the triangles, or 
equivalently $\phi^3$ graphs with the Ising spins placed at the vertices.

Let us first consider a closed, connected, planar graph $G$ consisting of vertices of degree $3$, 
and define the Ising model on the graph $G$: 
\[
Z_G (\beta) = \sum_{\{ \sigma \}} \prod_{<i,j>} e^{\beta  \sigma_i \sigma_j}\ , 
\label{eq:Isingpartfun}
\]
where $\beta$ is the inverse temperature, 
$\sigma_i$ being $\pm 1$ a spin located at a vertex $i$, 
$\sum_{\{ \sigma \}}$ a sum over all spin configurations 
and $ \prod_{<i,j>}$ a product with respect to all nearest-neighbor pairs of vertices. 
The Boulatov-Kazakov Ising model is given by a sum of (\ref{eq:Isingpartfun}) over all possible closed, connected, planar graphs: 
\[
F_{\text{BK}} (g,c) 
= \sum_G \frac{1}{|\text{Aut}(G)|} 
\left(
\left(
\frac{\sqrt{c}}{1-c^2}
\right)^{3/2} g 
\right)^{n(G)}  
Z_G (\beta)\ ,
\label{eq:BKgrandpartfun}
\]   
where $|\text{Aut}(G)|$ is the order of automorphism group of $G$, 
$n(G)$ the total number of vertices in $G$, 
$g$ essentially a weight for each vertex 
and an additional weight 
$(
\sqrt{c}/ (1-c^2))^{3/2}$ enters to make a connection with a matrix model introduced in due course.  
Here $c$ and $g$ are related to the inverse temperature and the dimensionless cosmological constant $\lambda$ 
such that $c=e^{-2\beta}$ and $g=e^{-\lambda}$, respectively. 
The $\phi^3$ graphs (or the dual triangulations) are all assumed to have spherical topology and 
the sum over these graphs is the lattice version of the integration over 2d (spherical) geometries, 
in this way coupling quantum gravity to the Ising model.

The sum (\ref{eq:BKgrandpartfun}) can be rewritten 
as a sum over the number of vertices:
\[
F_{\text{BK}} (g,c) 
= \sum_n 
\left(
\left(
\frac{\sqrt{c}}{1-c^2}
\right)^{3/2} g 
\right)^{n}  
Z_n (\beta)\ .
\label{eq:BKgrandpartfun2} 
\] 
We can view $n$ as proportional to the volume (the area A) of 2d spacetime, 
since in the dual graph $n$ is the number of triangles, which we all consider 
having the same area proportional to $\varepsilon^2$,  where $\varepsilon$ is the length of a link in the 
triangulations. Thus $A(n) \propto n \cdot \varepsilon^2$, and $Z_n (\beta)$ can be understood as 
the  partition function of the Ising model of a fixed (spacetime) volume, but dressed by quantum gravity 
since it contains effects coming from the sum over all possible spherical triangulations with a given  $n$. 
The power series (\ref{eq:BKgrandpartfun2}) is convergent since for $n \gg 1$ 
\[
Z_n (\beta) \propto (1/g_k(\beta))^n n^{\gamma_s-3} (1+\mathcal{O}(1/n))\ , 
\label{eq:largenpartfun}
\]
where $\gamma_s$ is a universal constant known as the string susceptibility exponent. 
A finite radius of convergence $g_k (\beta)$, sometimes also called a critical coupling constant, 
essentially defines the free energy per vertex in the thermodynamic limit:
\[
f(\beta) 
= - \frac{1}{\beta} \lim_{n \to \infty} \frac{1}{n} \log Z_n (\beta) 
= \frac{1}{\beta} \log \left[ 
\left(
\frac{\sqrt{c}}{1-c^2}
\right)^{3/2}
g_k (\beta)
\right]\ ,
\label{eq:bkfreeenergy}
\] 
and the critical coupling constant has been computed \cite{Kazakov:1986hu, Boulatov:1986sb}:
\[
g^2_k (\beta) 
= (\rho^2 - c^2) \frac{(1+\rho)(-1+2c+\rho)}{8\rho}
+2c \left( \frac{(1+\rho)(-1+2c+\rho)}{8\rho} \right)^2\ ,
\label{eq:gk}
\]
where $c=e^{-2\beta}$. 
In the low-temperature regime 
\[
\rho = - \sqrt{\frac{1-2c}{3}}\ ,
\label{eq:rholow}
\]
while in the high-temperature regime 
\[
\rho = 
- \frac{\left( 2c(1-c)\sqrt{1-2c} + c (2+c(c-4)) \right)^{1/3}}{2} 
- \frac{c}{2} \left( 1 + \frac{c}{\left( 2c(1-c)\sqrt{1-2c} + c (2+c(c-4)) \right)^{1/3}} \right)\ . 
\label{eq:rhohigh}
\]

\begin{figure}[h]
\centering
\includegraphics[width=3in]{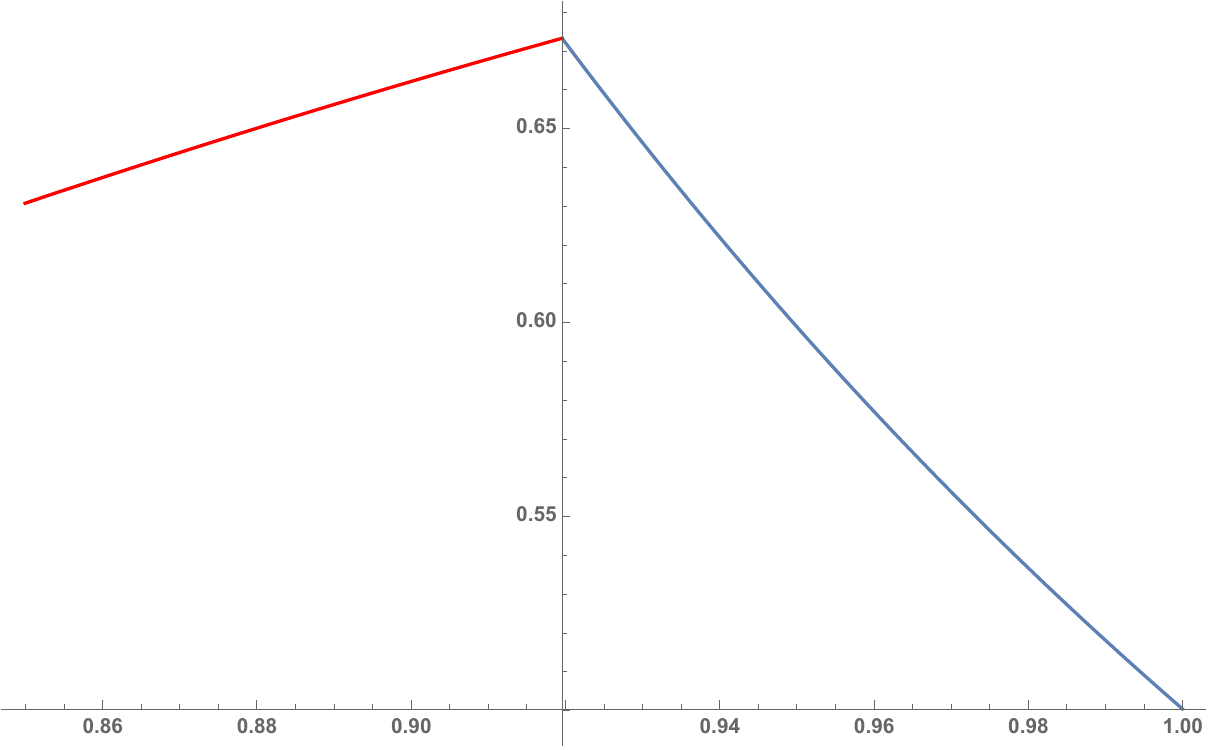}
\caption{A plot of the specific heat as a function of $\beta$. 
The curve has a cusp at $\beta_k$.    
}
\label{fig:specificheat}
\end{figure}

At the critical temperature $\beta^{-1}_k$ or equivalently 
\[
c_k := e^{-2\beta_k} = \frac{2\sqrt{7}-1}{27}\ , 
\label{eq:ck}
\]
the free energy and its first derivative are differentiable  functions, 
but the specific heat, a second derivative of the free energy, 
\[
C = - \beta^2 \frac{\partial^2}{\partial \beta^2} (\beta f (\beta))\ , 
\label{eq:specificheat}
\]
has a cusp (see Fig. \ref{fig:specificheat}).  
This is a signal of a third-order phase transition. 
As mentioned, introducing a magnetic field into the system, all critical exponents have been calculated: 
$\alpha=-1\ , \beta =1/2\ , \gamma=2\ , \delta=5\ , d \nu=3$ \cite{Boulatov:1986sb}, 
and they are different from the flat-space Onsager exponents.

\subsection{Continuum limit}
\label{sec:continuumlimit}
The power series (\ref{eq:BKgrandpartfun2}) become singular at $g=g_k$ 
and its singular behavior is characterized by the appearance of a fractional power 
in the expansion around $g_k$:
\[
Z(g,c) 
= c_0 + c_1 (g_k-g) + c_2 (g_k-g)^2 + c_{2-\gamma_s} (g_k-g)^{2-\gamma_s} + \cdots\ , 
\label{eq:expandz} 
\]
where $c_0$, $c_1$, $c_2$ and $c_{2-\gamma_s}$ are numerical constants and 
$\gamma_s$, string susceptibility exponent, quantifies the singularity. 
We have $\gamma_s=-1/2$ at $\beta \ne \beta_k$ while it changes at $\beta_k$ to $-1/3$. 
As mentioned this change is a result  of the back reaction of spins on the  geometries.  

The fractional power in the expansion implies that if one differentiates $Z(g,c)$ suitable times with respect to $g$, 
it diverges at $g=g_k$. As a result, the average number of vertices blows up when tuning $g$ to $g_k$: 
\[ 
\langle n \rangle = g \frac{\partial}{\partial g} \log Z(g,c) \biggl|_{\text{sing}} \sim \frac{1}{g_k-g}\ , 
\label{eq:averagen}
\]  
where ``sing'' means to pick up the singular part. 

Using this singular behavior, one can define the continuum limit. 
The large-$n$ asymptotic behavior of the partition function (\ref{eq:largenpartfun}) implies, 
for $n \gg 1$,  
\[
g^n Z_n (\beta) \propto (g/g_k)^n n^{\gamma_s-3} 
= e^{-(\lambda-\lambda_k)n} n^{\gamma_s-3}\ , 
\label{eq:contlim1}
\]
where $g_k=:e^{-\lambda_k}$. 
As can be understood from (\ref{eq:averagen}), this large-$n$ behavior becomes important if tuning $g$ to $g_k$. 
Therefore, introducing the lattice spacing $\varepsilon$ of triangulations one can, as mentioned above, define
the physical area $A$ and the renormalized cosmological constant $\Lambda$ by
\[
A=\varepsilon^2n\ , \ \ \ \Lambda = \frac{\lambda - \lambda_k}{\varepsilon^2}\ , 
\label{eq:areaandlambda}
\] 
and the continuum limit is obtained by tuning $g\to g_k$ (and correspondingly  $\varepsilon \to 0$) such that 
$\Lambda$ is kept fixed and  $A$ finite. 
Taking this continuum limit, one obtains
\[
g^n Z_n (\beta)  \propto e^{-\Lambda A} A^{\gamma_s-3}\ . 
\label{eq:contlim2} 
\] 
This quantity can be compared with the path-integral of the Liouville theory coupled to conformal field theories 
with  fixed area $A$. At $\beta\ne \beta_k$, it coincides with that of the Liouville theory  for pure gravity,
i.e. Liouville theory coupled to matter fields  with $c=0$. 
At the critical temperature, the spin fluctuations diverge, and as a result one obtains instead  
the behavior of the Liouville theory coupled to a $c=1/2$ conformal field theory.

\subsection{The matrix model representation}
\label{sec:twomatrixmodel} 
The so-called matrix models allow us to implement the sum over graphs via simple Gaussian integrals.
In the case of the Boulatov-Kazakov Ising model, the following two-matrix model plays that role \cite{Kazakov:1986hu, Boulatov:1986sb}:  
\[
Z_N (g,c) 
= \int D \psi_+ D\psi_-\ e^{-N\text{tr} V(\psi_+,\psi_-)}\ , 
\label{eq:BKmatrixmodel}
\] 
where $\psi_{\pm}$ are Hermitian $N$$\times$$N$ matrices, $D\psi_{\pm}$ the Haar measures on Hermitian matrices 
and the potential
\[
V(\psi_+,\psi_-) 
= \frac{1}{2} \psi^2_+ 
+ \frac{1}{2} \psi^2_- 
-c\psi_+\psi_- 
-\frac{g}{3} \left( \psi^3_+ + \psi^3_- \right)\ . 
\label{eq:v}
\]
The integral (\ref{eq:BKmatrixmodel}) is  defined formally as a power series with respect to $g$, 
and the coefficient to $g^n$  generates Feynman graphs with $n$ vertices of degree $3$. The vertices 
associated with $\psi_+^3$ can be thought of as having Ising spin  $\sigma=1$ and the vertices associated with
$\psi_-^3$ as having Ising spin  $\sigma=-1$. 
By Wick's theorem, the integral (\ref{eq:BKmatrixmodel}) implements the sum over all possible graphs 
with the nearest-neighbor spin interactions taken into account  properly if $c=e^{-2\beta}$. 
If one takes the matrix size $N$ to be large, one can suppress non-planar graphs in the sum. 
With this understanding, the Boulatov-Kazakov model can be defined by the matrix model as follows:   
\[
F_{\text{BK}} (g,c) := \lim_{N\to \infty} \frac{1}{N^2} \log \left( \frac{Z_N(g,c)}{Z_N(0,c)} \right)\ , 
\label{eq:defbymatrixmodel}
\]
where the logarithm is needed to single out connected graphs.

\section{A setup for cooling} 
\label{sec:asetupforcooling}

To reduce the critical temperature of the Ising model on $2$d DT down to the zero temperature 
and examine its critical behavior the following matrix model has been proposed \cite{Sato:2017ccb,Fuji:2011ce}: 
\[
I_N (g, c, \theta) 
&= \int D\varphi_+ D\varphi_- \
e^{ - N \text{tr}
U (\varphi_+, \varphi_-) }\ , 
\label{eq:ASTmatrixmodel}
\]
where $\varphi_{\pm}$ are Hermitian $N$$\times$$N$ matrices, 
$D\varphi_{\pm}$ are the Haar measures on Hermitian matrices and the potential is given by 
\[
U (\varphi_+, \varphi_-) 
= 
\frac{1}{\theta} 
\left(
\frac{1}{2} \varphi^2_+ + \frac{1}{2} \varphi^2_- 
-c\varphi_+ \varphi_-
- g \left( \varphi_+ + \varphi_- \right) 
-\frac{g}{3} \left( \varphi^3_+ + \varphi^3_- \right)
\right)\ .
\label{eq:u}
\]

\begin{figure}[h]
\centering
\includegraphics[width=6in]{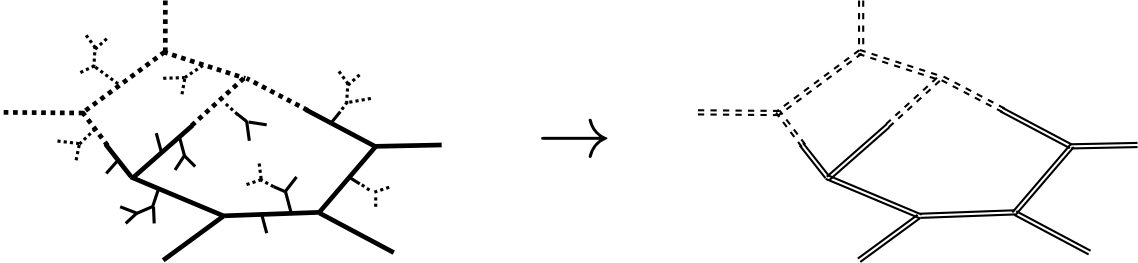}
\caption{The left figure: 
A typical planar graph generated by the potential (\ref{eq:u}) 
in which tree graphs attached to a skeleton graph.  
Each solid and dotted edges correspond to the propagators, 
$\langle \varphi_+ \varphi_+ \rangle_0$ 
and $\langle \varphi_- \varphi_- \rangle_0$, respectively; 
each half-solid and half-dotted edge corresponds to the propagator, 
$\langle \varphi_+ \varphi_- \rangle_0$ or $\langle \varphi_- \varphi_+ \rangle_0$.   
The right figure: A typical planar graph generated by the potential (\ref{eq:tildeu}) 
in which all tree graphs are integrated out. 
Each solid double-line and each dotted double-line correspond to the propagators, 
$\langle \tilde\varphi_+ \tilde\varphi_+ \rangle_0$ 
and $\langle \tilde\varphi_- \tilde\varphi_- \rangle_0$, respectively; 
each half-solid and half-dotted double-line corresponds to the propagator, 
$\langle \tilde\varphi_+ \tilde\varphi_- \rangle_0$ or $\langle \tilde\varphi_- \tilde\varphi_+ \rangle_0$.
}
\label{fig:integrateout}
\end{figure}

Perturbative expansions with respect to $g$ give Feynman graphs consisting of 
vertices of degree $1$ and $3$. A typical planar graph is a skeleton graph with tree graphs attached (see the LHS of Fig. \ref{fig:integrateout}). 
Here the skeleton graph means a planar graph consisting only of vertices of degree $3$. 
The parameter $\theta$ is a loop-counting parameter meaning that if $\theta \ll 1$, loops in Feynman graphs are suppressed and tree structures become dominant.      
This kind of modification was first introduced in the context of a one-matrix model in order 
to obtain GCDT from DT and it defines a new continuum limit of one-matrix models \cite{Ambjorn:2008gk,Ambjorn:2008jf}. The detailed disentanglement of the model in tree- and skeleton graphs
and possible scaling limits associated with this was studied in \cite{Ambjorn:2014bga}.

The two-matrix model (\ref{eq:ASTmatrixmodel}) defines a slightly ``modified'' Ising model on $2$d DT: 
\[
F_{\text{ST}} (g,c,\theta) 
&:= \lim_{N\to \infty} \frac{1}{N^2} \log \left( \frac{I_N(g,c,\theta)}{I_N(0,c,\theta)} \right) \notag \\
&= \sum_G \frac{1}{|\text{Aut}(G)|} 
\left( \left( \frac{\sqrt{c}}{1-c^2}  \right)^{1/2} \frac{g}{\sqrt{\theta}} \right)^{n_1(G)}
\left( \left( \frac{\sqrt{c}}{1-c^2} \right)^{3/2} g \sqrt{\theta} \right)^{n_3(G)}
Z_G (\beta)
\ , 
\label{eq:defbymatrixmodel2}
\] 
where $G$ denotes a closed, connected, planar graph generated by the matrix model with the potential (\ref{eq:u}), 
and $n_1$ and $n_3$ the total number of vertices of degree $1$ and $3$, respectively. 
The critical temperature of this system becomes a function of $\theta$,    
and as shown in \cite{Sato:2017ccb}, when $\theta \ne 0$, the critical behavior is nothing but that of the Boulatov-Kazakov Ising model, 
but when tuning $\theta \to 0$ the critical temperature reaches  zero temperature at which the tree structures become dominant, 
resulting in a continuum theory different from the Liouville theory coupled to a conformal matter with $c=1/2$.

It is useful to classify vertices in a graph $G$ into two kinds: skeleton vertices and others.  
Let us pick up a vertex of degree $3$ and label the three links emanating from the vertex, say $1$, $2$ and $3$ in a clockwise manner. 
When moving from that vertex to other vertices via links, if one can find a path coming back to the first vertex whichever link $1$, $2$ or $3$, one starts with,  
the vertex picked up is called a skeleton vertex. 
Since $\theta$ is a loop-counting parameter, the number of skeleton vertices is supposed to be controlled by $\theta$. 
Concerning (\ref{eq:defbymatrixmodel2}), if we implement the redefinition
\[
n_t(G) := n_1(G) + n_3 (G)\ , \ \ \ 
n_s(G) := n_3 (G) - n_1 (G)\ , 
\label{eq:newns}
\] 
where $n_t(G)$ and $n_s(G)$ are the total number of vertices and the number of skeleton vertices in a given graph $G$, 
we obtain 
\[
F_{\text{ST}} (g,c,\theta) 
= 
\sum_G \frac{1}{|\text{Aut}(G)|}\  
g^{n_t(G)}_t g^{n_s(G)}_s 
Z_G (\beta)\ , 
\label{eq:defbymatrixmodel3}
\]
where 
\[
g_t = \frac{\sqrt{c}}{1-c^2}\ g\ , \ \ \ g_s = \left(  \frac{\sqrt{c}}{1-c^2} \right)^{1/2} \sqrt{\theta}\ .
\label{eq:gtandgs}
\]
From (\ref{eq:defbymatrixmodel3}) and (\ref{eq:gtandgs}), it is indeed $\theta$ that controls the number of skeleton vertices, 
while the total number of vertices is controlled by $g$.

\subsection{Relation to the Boulatov-Kazakov Ising model}
\label{sec:relationtotheboulatovkazakovisingmodel}

As shown in \cite{Sato:2017ccb}\footnote{In the one-matrix model case, a similar transformation has been considered in \cite{Ambjorn:2014bga}.}, 
one can map our matrix model defined by (\ref{eq:defbymatrixmodel2}) to the matrix model for the Boulatov-Kazakov Ising model (\ref{eq:BKmatrixmodel}). 
This relation turns out to be useful for understanding the physics of our Ising model.

Starting from the matrix model (\ref{eq:defbymatrixmodel2}) and changing variables
\[
\varphi_{\pm} =  \tilde \varphi_{\pm} + Z_{\text{tree}} (g,c)\ , \ \ \ 
\text{with}\ \ \ 
Z_{\text{tree}}(g,c) = \frac{1-c-\sqrt{(1-c)^2 -4g^2}}{2g}\ ,
\label{eq:tildevarhpipm}
\] 
the integral (\ref{eq:defbymatrixmodel2}) becomes 
\[
I_N (g,c,\theta) = e^{N^2 F_{\text{cons}} (g,c,\theta)} \int D\tilde\varphi_+ D\tilde\varphi_-\ e^{-N\text{tr} \tilde U (\tilde\varphi_+, \tilde\varphi_-)}\ ,
\label{eq:partitionfunction2}
\] 
where
\[
\tilde U (\tilde\varphi_+,\tilde\varphi_-)  
&= \frac{1}{\theta} 
\left(  
\frac{1-2gZ_{\text{tree}} (g,c)}{2} \left( \tilde\varphi^2_+ + \tilde\varphi^2_- \right)
-c \tilde\varphi_+\tilde\varphi_-
-\frac{g}{3} \left( \tilde\varphi^3_+ + \tilde\varphi^3_- \right)
\right)\ ,
\label{eq:tildeu} \\
F_{\text{cons}}(g,c,\theta)&=
\frac{1}{6 \theta g^2} 
\biggl(
1-c^3 - \sqrt{ (1-c)^2 -4g^2 } 
-g^2 \left( 6-4  \sqrt{ (1-c)^2 -4g^2 } \right) \notag \\ 
&\ \ \ -c \left(3 - 6g^2 -2  \sqrt{ (1-c)^2 -4g^2 } \right) 
+ c^2 \left(3 -   \sqrt{ (1-c)^2 -4g^2 } \right) 
\biggl)\ . 
\label{eq:a}
\]
Here $Z_{\text{tree}}$ is the sum of all connected planar, rooted tree graphs with a spin placed at each vertex of degree $1$ and $3$ 
as well as the sum over all spin configurations \cite{Sato:2017ccb}.

Through this transformation, the linear terms in (\ref{eq:u}) are integrated out 
and a typical Feynman graph is depicted in the RHS of Fig. \ref{fig:integrateout}, 
which is a skeleton graph with dressed edges.  
The non-canonical quadratic terms in (\ref{eq:tildeu}) contribute to the dressed edges through the dressed propagators
\[
\left\langle \tilde\varphi_{\pm} \tilde\varphi_{\pm}  \right\rangle_0 
&=  \frac{1-2gZ_{\text{tree}}}{ (1-2gZ_{\text{tree}})^2 -c^2 }\ \langle \varphi_{\pm}\varphi_{\pm} \rangle_0 \ , \ \ \ \text{with}\ \ \  \langle \varphi_{\pm}\varphi_{\pm} \rangle_0 = \frac{\theta}{N}\ ,
\label{eq:prophomo} \\
\left\langle \tilde\varphi_{\pm} \tilde\varphi_{\mp}  \right\rangle_0 
&=  \frac{1}{ (1-2gZ_{\text{tree}})^2 -c^2 }\ \langle \varphi_{\pm}\varphi_{\mp} \rangle_0 \ , \ \ \ \text{with}\ \ \  \langle \varphi_{\pm}\varphi_{\mp} \rangle_0 = \frac{\theta c}{N}\ ,
\label{eq:prophetero} 
\]
where the indices of the matrices have been omitted. 
The dressed propagators (\ref{eq:prophomo}) and (\ref{eq:prophetero}) 
can be obtained by summing all possible tree outgrowths from the canonical propagators, 
$\langle \varphi_{\pm}\varphi_{\pm} \rangle_0$ and $\langle \varphi_{\pm}\varphi_{\mp} \rangle_0 $ \cite{Sato:2017ccb}.

Rescaling the new variables 
\[
\tilde \varphi_{\pm} = \sqrt{ \frac{\theta}{1-2g Z_{\text{tree}}(g,c)}}\ \psi_{\pm}\ , 
\label{eq:psipm}
\] 
the integral (\ref{eq:partitionfunction2}) becomes
\[
I_N (g,c,\theta)
= \left( \frac{\theta}{c+\sqrt{(1-c)^2-4g^2}} \right)^{N^2} e^{N^2 F_{\text{cons}} (g,c,\theta)}\  
Z_N (c_{\text{BK}}, g_{\text{BK}})\ , 
\label{eq:partitionfunction3}
\] 
where $Z_N$ is nothing but the matrix model for the Boulatov-Kazakov Ising model, 
with the coupling constants $(g,c)$ substituted by the ``Boulatov-Kazakov'' coupling constants  
 $(c_{\text{BK}} ,g_{\text{BK}})$  defined by 
\[
c_{\text{BK}} 
&= \frac{c}{1-2gZ_{\text{tree}}(g,c)}
= \frac{c}{c+\sqrt{(1-c)^2 - 4g^2}}\ , \label{eq:cbk} \\
g_{\text{BK}} 
&=\frac{\theta^{1/2} g}{(1-2gZ_{\text{tree}}(g,c))^{3/2}}
=\frac{\theta^{1/2} g}{( c+ \sqrt{(1-c)^2 - 4g^2} )^{3/2}}\ . 
\label{eq:gbk}
\] 

As a result, we obtain 
\[
F_{\text{ST}} (g,c,\theta) 
= F_{\text{BK}} (g_{\text{BK}}   ,    c_{\text{BK}}   ) 
+ F_{\text{tree}} (g,c) 
+ F_{\text{cons}} (g,c,\theta)\ , 
\label{eq:relation}
\]
where 
\[
F_{\text{tree}} (g,c) 
= - \log \left[ c + \sqrt{(1-c)^2 - 4g^2} \right]\ . 
\label{eq:ftree}
\]

\begin{figure}[h]
\centering
\includegraphics[width=3in]{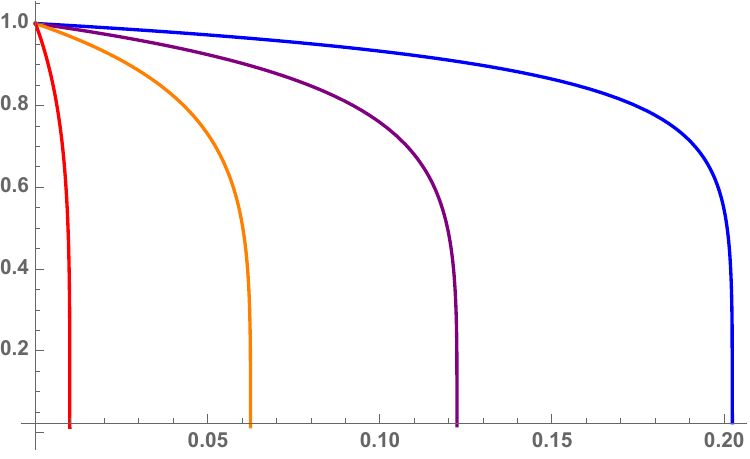}
\caption{Plots of $\sqrt{z}$ as a function of $g^2$ for a given value of $c$: 
The blue, purple, orange and red curves are those for $c=0.1, 0.3, 0.5, 0.8$, respectively.      
}
\label{fig:spinrenormalization}
\end{figure}

Through the change of matrix variables, the inverse temperature in our original system,  
$\beta=-\log [c]/2$, 
changes to 
\[
\beta_{\text{BK}} 
=- \frac{1}{2} \log \left[c_{\text{BK}}(g,c) \right] 
= \beta 
\left(
1 - \frac{\log \left[  c + \sqrt{(1-c)^2 - 4g^2} \right]}{\log \left[ c \right]} 
\right)\ .  
\label{eq:betabk}
\]
We can think of this change of temperature as a change of spins, i.e.  
the nearest-neighbor spin interaction changes if integrating trees out: 
\[
-\beta \sum_{<i,j> \in v(G)} \sigma_i \sigma_j \ \ \ 
\Rightarrow 
\ \ \ 
-\beta_{\text{BK}} \sum_{<i,j> \in v(G_s)} \sigma_i \sigma_j\ , 
\label{eq:changeofnearestneighbor}
\]
where $<i,j> \in v(G)$ ($<i,j> \in v(G_s)$) denotes a pair of nearest-neighbor vertices $i$ and $j$ in a set of vertices in a graph $G$ ($G_s$, a skeleton graph), 
and then we can define effective spins $\tilde \sigma_i (g,c)$'s by the following equation:
\[
-\beta_{\text{BK}} \sum_{<i,j> \in v(G_s)} \sigma_i \sigma_j 
=:
-\beta \sum_{<i,j> \in v(G_s)} \tilde \sigma_i (g,c) \tilde \sigma_j (g,c)\ ,
\label{eq:effectivespin1}
\]
where
\[
\tilde \sigma_i (g,c) 
=
\left(
1 - \frac{\log \left[ c + \sqrt{(1-c)^2 - 4g^2} \right]}{\log \left[ c \right]} 
\right)^{1/2} \sigma_i 
=: \sqrt{z} \sigma_i\ .
\label{eq:effectivespin2}
\]
One can show that $0\le \sqrt{z} \le 1$ if $g \le (1-c)/2$ for a given $c$. 
A qualitative behavior of the ``spin renormalization'' $\sqrt{z}$ can be seen in Fig. \ref{fig:spinrenormalization}. 

At  zero temperature $c=0$ we have $\sqrt{z}=1$ for $g \le 1/2$. 
On the other hand, at  high temperature $c=1$ we have  $g=0$ and $\sqrt{z}=0$.  
As one can see from Fig. \ref{fig:spinrenormalization}, the effective spins behave like ordinary spins in the low-temperature regime 
since the spin renormalization is not that sensitive to $g$ and its value is almost $1$. 
However, in the high-temperature regime the spin renormalization is very sensitive to $g$ 
and rapidly goes down when increasing $g$ up to its critical value $(1-c)/2$. 
This means that at the high temperature, the effective spins strongly ``feel'' the existence 
of tree structures (already integrated out) which weaken the effective spins.

\section{Criticality}
\label{sec:criticality}

In this section, we study the critical behavior of our Ising model defined in terms of the matrix model (\ref{eq:ASTmatrixmodel}), focusing especially on the zero temperature regime.

The critical curve on which spin fluctuations diverge is determined by the set of equations \cite{Sato:2017ccb}:
\[
1-4g^2_k 
&= \frac{2^{2/3}}{5^{1/3}} (\theta g^2_k)^{1/3} \left( 1 + \frac{27}{2 \times 10^{1/3}} ( \theta g^2_k)^{1/3}  \right)\ , \label{eq:critical1} \\
c_k &= \frac{1}{10^{1/3}} (\theta g^2_k)^{1/3}\ .
\label{eq:critical2}
\]
Removing $\theta$ from (\ref{eq:critical1}) and (\ref{eq:critical2}), 
we find the critical curve:
\[
c_k=\frac{2\sqrt{7-27g^2_k}-1}{27}\ .
\label{eq:criticalBK}
\]
Inserting (\ref{eq:criticalBK}) into (\ref{eq:cbk}), we recover the critical point obtained by \cite{Boulatov:1986sb}: 
\[
(c_{\text{BK}})_k = \frac{2\sqrt{7}-1}{27}\ , \ \ \ 
(g_{\text{BK}})_k = \frac{\sqrt{10}}{( 1+ 2\sqrt{7} )^{3/2}}\ . 
\label{eq:bkcriticalpoint}
\]
Therefore, on the critical curve (\ref{eq:criticalBK}) except at the  endpoint $\theta =0$ 
we should observe the same criticality as that of the Boulatov-Kakakov Ising model (\ref{eq:bkcriticalpoint}). 
In this sense, let us call the curve (\ref{eq:criticalBK}) the Boulatov-Kazakov critical curve. The reason the 
the critical point of the Boulatov-Kazakov Ising model has been replaced by a curve is that our model has 
the additional parameter $\theta$ and we can use this parameter in a 
parametric representation of the Boulatov-Kazakov critical curve.  
From (\ref{eq:critical1}) and (\ref{eq:critical2}), we can determine $c_k$ and $g_k$ as functions of $\theta$ \cite{Sato:2017ccb}:
\[
c_k (\theta) 
&= \frac{\theta^{1/3}}{10^{1/3}} 
\left( 
- \frac{9 \theta^{2/3}}{4 \times 10^{2/3}} 
+ \frac{3^{1/3} \theta^{1/3} (243\theta - 80) + H^2}{4 \times 30^{2/3} H}
\right)\ , \label{eq:criticalc} \\
g_k (\theta) 
&= \left( 
- \frac{9 \theta^{2/3}}{4 \times 10^{2/3}} 
+ \frac{3^{1/3} \theta^{1/3} (243\theta - 80) + H^2}{4 \times 30^{2/3} H}
\right)^{3/2}\ , \label{eq:criticalg}
\]
where 
\[
H = 
\left[ 
81 (40-81\theta)\theta 
+ 80 \left( 90 + \sqrt{ 8100 + 3 (2510 - 5103 \theta)\theta } \right)
\right]^{1/3}\ . 
\label{eq:h}
\]

In addition, we have another critical curve for dominant trees determined by the condition that 
the average number of vertices in the dressed propagators (\ref{eq:prophomo}) and (\ref{eq:prophetero}), 
\[
g\frac{\partial}{\partial g} \log \langle \tilde \varphi_{\pm} \tilde \varphi_{\pm} \rangle_0\ , \ \ \ 
g\frac{\partial}{\partial g} \log \langle \tilde \varphi_{\pm} \tilde \varphi_{\mp} \rangle_0\ ,  
\label{eq:nprop}
\]
diverge, which yields  
\[
c=1-2 g\ . 
\label{eq:criticaltree}
\]
Inserting (\ref{eq:criticaltree}) into (\ref{eq:cbk}) and (\ref{eq:gbk}), 
we have 
\[
c_{\text{BK}} = 1\ , \ \ \ 
g_{\text{BK}} = \frac{g\sqrt{\theta}}{(1-2g)^{3/2}} 
\ .\label{eq:cgbktree}
\]
From (\ref{eq:gk}),  
\[
(g_{\text{BK}})_k (c_{\text{BK}}=1)=0\ . 
\label{eq:gbktree} 
\]
Therefore, we conclude that $\theta=0$ on the critical curve for dominant trees.

We can compute the free energy per vertex on the critical curve for dominant trees. 
Let us rewrite (\ref{eq:defbymatrixmodel3}) as  
\[
F_{\text{ST}} (g,c,\theta) 
= 
\sum_{n_t,n_s} 
g^{n_t}_t g^{n_s}_s 
Z_{n_t,n_s} (\beta)
=
\sum_{n_s} g^{n_s}_s 
\sum_{n_t} g^{n_t}_t Z_{n_t,n_s} (\beta)
\ . 
\label{eq:defbymatrixmodel4}
\]
For a given finite $n_s$, the radius of convergence for the power series of $g_t$ yields 
\[
(g_t)_k (\beta) = 
\frac{g_* \sqrt{c}}{1-c^2} 
= \frac{\sqrt{c}}{2(1+c)}\ , 
\label{eq:gtk}
\]
where we have used (\ref{eq:gtandgs}) and $g_*$ is a solution to (\ref{eq:criticaltree}). 
The free energy per vertex on the critical curve for dominant trees is 
\[
f_{\text{tree}} (\beta) 
= - \frac{1}{\beta} \lim_{n_t \to \infty} \frac{1}{n_t} \log Z_{n_t,n_s} (\beta) \notag 
= \frac{1}{\beta} \log \left[ (g_t)_k (\beta) \right] \notag 
=-\frac{1}{\beta} \log \left( 4 \cosh [\beta] \right)\ .
\] 
This is essentially the free energy per vertex of the classical $1$d spin chain, or that of the Ising model on branched polymers \cite{Ambjorn:1992rp}. 
Thus, along the critical curve for dominant trees, the system is magnetized only at  zero temperature.

\begin{figure}[h]
\centering
\includegraphics[width=3in]{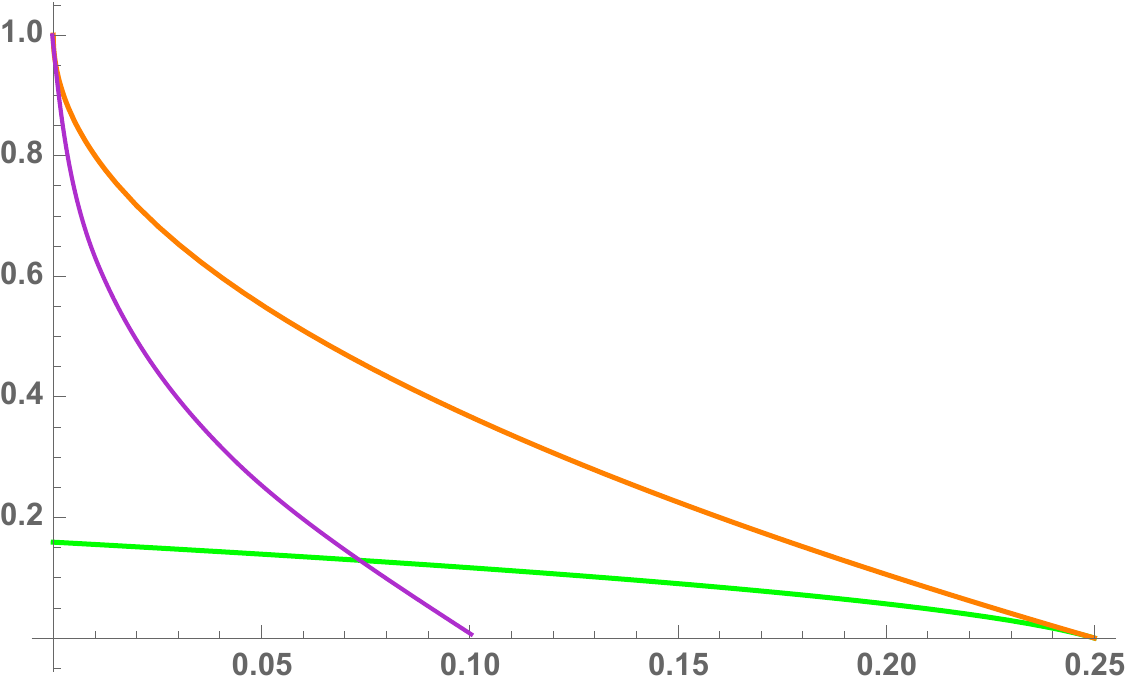}
\caption{The phase diagram: 
The vertical axis is $c$ and the horizontal one is $g^2$. 
The green curve is the Boulatov-Kazakov critical curve that separates the high- and low-temperature phases and the orange curve is the critical curve for dominant trees. 
The purple curve is a schematic critical curve for the Liouville gravity for a fixed $\theta$.          
}
\label{fig:phasediagram}
\end{figure}  
Fig. \ref{fig:phasediagram} is the phase diagram. 
For a fixed $\theta> 0$ we have a well defined map given by eqs.\ (\ref{eq:cbk}) (\ref{eq:gbk}) between 
$(g,c)$ and $(g_{\text{BK}},c_{\text{BK}})$. In the Boulatov-Kazakov model we have a critical curve: for a given temperature $\beta_{\text{BK}}^{-1}$, i.e.\
a given $c_{\text{BK}} = e^{-2\beta_{\text{BK}}} $, there is a corresponding critical $g_{\text{BK}}$ where the
continuum limit is that of Liouville gravity. This defines the critical curve in the $(g_{\text{BK}},c_{\text{BK}})$-plane. 
One point on this curve is special, namely the point given by eq.\ (\ref{eq:bkcriticalpoint}), 
where the Liouville gravity theory changes from pure 2d gravity to 2d gravity coupled to
a $c=1/2$ conformal theory. For a fixed value of $\theta >0$ one can draw the corresponding curve in the 
$(g,c)$-plane. This is the the purple curve in Fig. \ref{fig:phasediagram}. For a fixed $\theta$ this is where
one can take the continuum limit in our modified model. It 
crosses the Boulatov-Kazakov critical curve where the Ising spins become critical.   For each $\theta$ one 
has a different curve. 
The vertical axis in the phase diagram corresponds to the case with $\theta$ being $\infty$, 
and with decreasing $\theta$ the curves gradually move to the right in such a way that they 
share the point $(g^2,c)=(0,1)$ and asymptotically approach the critical curve for dominant trees 
on which $\theta=0$. To reach the zero-temperature critical point at which two kinds of critical lines meet, 
one has to tune $\theta$ to $0$.

\subsection{Zero-temperature criticality}
\label{sec:zerotemperaturecriticality}

The physics around the zero-temperature critical point is very sensitive to the way we approach the point. 
In the following we show several examples.    

One way to approach the zero-temperature critical point,  
proposed in \cite{Sato:2017ccb} is 
\[
g(\theta) = g_k(\theta) (1-\varepsilon^2 \Lambda)\ , \ \ \ 
c(\theta) = c_k(\theta)\ , \ \ \ 
\theta = \varepsilon^3 \Theta\ , 
\label{eq:scaling1}
\]
where $\Lambda$ and $\Theta$ are dimension-full coupling constants. 
$g(\theta)$ is chosen to be slightly away from the Boulatov-Kazakov critical curve
in the spirit of (\ref{eq:areaandlambda}), allowing us to interpret $\Lambda$ as a cosmological constant, 
while the temperature is chosen such that tuning $g(\theta) \to g_k(\theta)$ for fixed $\theta$ we 
would obtain Liouville gravity coupled to conformal matter with $c=1/2$. However, here we now
take the limit where $\varepsilon \to 0$, i.e.\ $\theta$ is scaled to 0 at the same time as $g(\theta) \to g_k(\theta)$. 
This  limit defines a continuum theory around the zero-temperature critical point that 
is described by a continuum two-matrix model \cite{Sato:2017ccb}.    
To analyze this limit (\ref{eq:scaling1}), let us introduce 
\[
\frac{c_{\text{BK}}}{(c_{\text{BK}})_k} 
&= \frac{c}{c_k} \left( \frac{c_k + \sqrt{(1-c_k)^2 - 4g^2_k}}{c+\sqrt{(1-c)^2 -4g^2}} \right)\ , \label{eq:rel1} \\
\frac{g_{\text{BK}}}{(g_{\text{BK}})_k} 
&= \frac{g}{g_k} \left( \frac{c_k + \sqrt{(1-c_k)^2 - 4g^2_k}}{c+\sqrt{(1-c)^2 -4g^2}} \right)^{3/2}\ , \label{eq:rel2}
\] 
which have been obtained from (\ref{eq:cbk}) and (\ref{eq:gbk}). 
Plugging (\ref{eq:scaling1}) into (\ref{eq:rel1}) and (\ref{eq:rel2}), one obtains at the small-$\varepsilon$ limit 
\[
\frac{c_{\text{BK}}}{(c_{\text{BK}})_k} 
&=  \frac{1+2\sqrt{7}}{1+2\sqrt{ 7 + 5^{2/3} \frac{\Lambda}{\Theta^{2/3}} }} \ , \label{eq:rel1a} \\
\frac{g_{\text{BK}}}{(g_{\text{BK}})_k} 
&=  \left( \frac{1+2\sqrt{7}}{1+2\sqrt{ 7 + 5^{2/3} \frac{\Lambda}{\Theta^{2/3}} }}  \right)^{3/2}\ . \label{eq:rel2a}
\]
Therefore, with the continuum limit (\ref{eq:scaling1}), 
spins on skeleton graphs cannot be critical 
and the criticality is governed by that of trees.

Next we introduce a continuum limit such that spins, tree graphs and skeleton graphs are all critical, 
which can be realized by the following choice: 
\[
g(\theta) = g_k(\theta) (1-\varepsilon^2 \Lambda)\ , \ \ \ 
c(\theta) = c_k(\theta)\ , \ \ \ 
\theta = \varepsilon^a \Theta_a\ , \ \ \ \text{with} \ \ \  0<a<3\ ,  
\label{eq:scaling2}
\]
where $\Theta_a$ is a dimension-full coupling constant. This $a$-dependent continuum limit has been first introduced in the context of one-matrix model \cite{Ambjorn:2014bga}.  
Inserting (\ref{eq:scaling2}) into (\ref{eq:rel1}) and (\ref{eq:rel2}), one obtains 
\[
\frac{c_{\text{BK}}}{(c_{\text{BK}})_k} 
&= 1 - \frac{5^{2/3}}{14 + \sqrt{7}} \frac{\Lambda}{\Theta^{2/3}_a} \varepsilon^{2-\frac{2}{3} a } + \cdots \ , \label{eq:rel1b} \\
\frac{g_{\text{BK}}}{(g_{\text{BK}})_k} 
&=   1 - \frac{3\times 5^{2/3}}{2(14 + \sqrt{7})} \frac{\Lambda}{\Theta^{2/3}_a} \varepsilon^{2-\frac{2}{3} a } + \cdots  \ . \label{eq:rel2b}
\] 
This means that if $0<a<3$, spins, tree graphs and skeleton graphs are all critical. 
Based on the parametrization (\ref{eq:scaling2}), we can show 
\[
n_t \sim \frac{1}{\varepsilon^2}\ , \ \ \ 
n_p \sim \frac{1}{\varepsilon^{2a/3}}\ , \ \ \ 
n_s \sim \frac{1}{\varepsilon^{2-2a/3}}\ , 
\label{eq:adependentn}
\]
where $n_t$, $n_p$ and $n_s$ are the total number of vertices, the number of vertices in the dressed propagators 
and the number of skeleton vertices.

\subsection{New critical exponents}

Let us first consider the approach to zero temperature given by eq.\ \rf{eq:scaling1}. 
Recall that the whole critical line of Fig.\ \ref{fig:phasediagram} is mapped to the single critical point 
$((g_{\rm BK})_k, (c_{\rm BK})_k)$ for the Boulatov-Kazakov Ising model on skeleton graphs. Similarly, the 
line defined by \rf{eq:scaling1} is (for small $\varepsilon$) mapped to a single point $(g_{\rm BK}, c_{\rm BK})$
given by eqs.\ \rf{eq:rel1a} and \rf{eq:rel2a}. This point has a finite distance (not necessarily small) to 
$((g_{\rm BK})_k, (c_{\rm BK})_k)$ and the Ising spins on the (finite number of) skeleton 
vertices are uncorrelated. On the other hand, for the tree graph related to a given link (the dressed propagators), 
the number of 
vertices is of the order $1/\vp^2$ while the temperature $\beta^{-1}$ is determined by $c = e^{-2\beta}$ and
$c \propto \theta^{1/3} \propto \vp$. Since the tree graph has Hausdorff dimension 2 and thus 
linear extension of order $1/\vp$, and since spin correlations in the tree behave essentially like 
on a linear chain, the correlation length is $\xi(\beta) = 1/c$. Thus we see that a given 
tree is essentially magnetized\footnote{By magnetized, we mean that the spins for a given configuration 
will essentially be alligned. However, since the volume is finite ($1/\vp^2$ for the tree) the ensemble 
average will have zero magnetization, as is well know for the linear spin chain at non-zero temperature.} 
A given spin configuration thus looks amazingly like a real, unmagnetized ferromagnetic material: it consists 
of a number of essentially magnetized regions (the dressed propagators), 
but the orientation of these is quite random, and the total 
magnetization of the piece of material is close to zero. However, we should stress that while there is 
this resemblance with a ferromagnet for a single configuration, the statistical properties are quite
different, since a tree never is magnetized for non-zero temperature. The magnetic properties are 
like those of a one-dimensional Ising chain as we have shown explicitly. 

Let us now turn to the more non-trivial scaling given by eq.\ \rf{eq:scaling2}, and characterized by the 
parameter $a$ between 0 and 3. For a given dressed propagator the situation is as before: the size of a 
typical tree associated with a dressed propagator is according to \rf{eq:adependentn} $n_p \sim 
1/\vp^{2a/3}$ and $c(\vp) \sim \vp^{a/3}$, i.e.\ the correlation length $\xi(\vp) \sim 1/\vp^{a/3}$. Thus
a tree in a given configuration is essentially magnetized. However, according to \rf{eq:adependentn}  we have now
a divergent number of skeleton vertices for $\vp \to 0$ and the corresponding Boulatov-Kazakov coupling constants
$(g_{\rm BK}(\vp), c_{\rm BK}(\vp))$ now approach $((g_{\rm BK})_k, (c_{\rm BK})_k)$ for $\vp \to 0$. The 
critical spin properties in such an approach is governed by two factors. The first one is that 
$c_{\rm BK}(\vp)\neq (c_{\rm BK})_k$. Thus, even if we had an infinite number of skeleton vertices, we 
would only have a finite spin correlation length
\beq\label{ja1}
\xi(\Delta c) \sim \frac{1}{|\Delta c|^\nu}\ , \quad \Delta c = (c_{\rm BK})_k- c_{\rm BK}\ .
\eeq
and approaching the critical point from the low temperature phase as we are doing, we would have a
magnetization per skeleton vertex
\beq\label{ja2}
m (\Delta c) \sim |\Delta c|^\beta\ ,
\eeq
where the critical exponent $\beta$ should not be confused with the inverse temperature also called $\beta$.
This magnetization per skeleton vertex would be present if we had an infinite volume, i.e.\ 
if we for the given $c_{\rm BK}(\vp)$, instead of  $g_{\rm BK}(\vp)$ had chosen the critical $(g_{\rm BK})_k$ corresponding 
to $c_{\rm BK}(\vp)$. However, we have a $g_{\rm BK}(\vp) \neq (g_{\rm BK})_k$, and thus we have a finite number of 
skeleton vertices $n_s \sim 1/\vp^{2-2\alpha/3}$, i.e.\ a finite volume $V_s$. In a finite volume we cannot determine 
the critical point $(c_{\rm BK})_k$ but we can determine the $c_{\rm BK}$ where the correlation length $\xi$ becomes 
of the order of the size of the system, or expressed in terms of the volume $V_s$: $\xi^d = V_s$, where 
$d$ is the dimension of the system\footnote{For a regular lattice the meaning of $d$ is clear, but for DT lattices 
the meaning is less clear. However, we will not need a precise definition since only the combination $\nu d$
enters in the discussion, and if hyper-scaling is valid then $\nu d$ is related to the critical exponent 
for the specific heat $\alpha = 2/\nu d$.}.  According to 
\rf{ja1} this corresponds to a $\Delta c$ of the order
\beq\label{ja3}
\Delta c = \frac{1}{V_s^{1/d \nu}}\ ,\qquad d \nu = 3\quad {\rm for~the~Boulatov-Kazakov~model}.
\eeq
We  now have  two $\Delta c$, which are functions of $\vp$:
\beq\label{ja4}
\Delta_1 c =  (c_{\rm BK})_k- c_{\rm BK}(\vp) \sim \vp^{2-2a/3} \sim \frac{1}{V_s (\vp)}\ ,  \quad {\rm and}\quad
\Delta_2 c = \left(\frac{1}{V_s(\vp)}\right)^{1/d \nu} =\left( \frac{1}{V_s(\vp)}\right)^{1/3}\ .
\eeq
Clearly $\Delta_1 c$ is irrelevant for the way we have chosen to approach zero temperature. 
Inserting into \rf{ja2} we obtain
\beq\label{ja5}
m (\vp)\sim V_s^{-\beta/d \nu} = V_s^{-1/6}\ ,
\eeq
which is just the standard result for the Boulatov-Kazakov model. However, viewed from outside where we 
do not insist in resolving  the graphs in dressed propagators and skeleton graphs, the dressed 
propagators do not in average contribute to the total spin, since this is already included in the mapping 
to the Boulatov-Kazakov model as emphasized in eq.\ \rf{eq:changeofnearestneighbor}. 
If we denote the number of vertices in the complete  graph by $V$,
we can write magnetization per vertex in the complete graph
 \beq\label{ja6}
m' (\vp)\sim  m(\vp) \;\frac{V_s}{V} \sim  \frac{1}{V^{(1+ 5a/3)/6}}\ .
\eeq
This is a new kind of critical behavior which interpolates between the Boulatov-Kazakov model, $a=0$, and 
the GCDT model $a=3$. However, in order to be able to identify the scaling behavior  with a critical exponent 
$\tilde{\beta}$, like in eq.\ \rf{ja2}, we have to write it in the form $V^{-\tilde{\beta}/\tilde{\nu} \tilde{d}}$, and it is
unclear how to think about exponents $\tilde{\nu}$ and $\tilde{d}$ in our model. Hyper-scaling usually 
links $\nu \, d$ to the exponent $\alpha$ for the specific heat by $\alpha = 2 - \nu d$. However, in our 
model there is no natural definition of the dimension $d$ valid at all scales: the trees have a 
fractal dimension that is  different from fractal dimension of the skeleton graph, so it seems unlikely that
such a hyper-scaling relation exists.


\section{Discussion}
\label{sec:discussion}

The DT graphs used to regularize 2d quantum gravity have generic fractal properties, among those
that the fractal dimension of the graphs is four \cite{hausdorff}. The generalized causal triangulations 
is another wide set of graphs, characterized by the property that the (graph)-distance from a 
vertex only has a finite number of local maxima, even for infinite graphs. The GCDT graphs
have fractal dimension two.
The coupling of matter to the ensembles of graphs is reasonably well understood in the DT case. 
If the matter system becomes critical for a certain choice of coupling constants, it defines a conformal field theory 
coupled to 2d gravity and both the critical properties of that matter system and the fractal properties of 
the ensemble of graphs change.

Coupling of matter theories to GCDT graphs are much less studied, but the interaction
between graphs and matter does not seem to change the critical properties of graphs or of the matter 
systems (if they have critical couplings). 
This is in agreement with the general expectation that   changes of the critical properties 
are caused by infinitely many baby universes. 
When coupling the Ising model to GCDT for instance, Onsager's critical exponents are expected to be recovered 
and the fractal dimension is expected to  be two in the continuum limit characterized by a finite number of baby universes. 
However, so far one has not been able to solve the model analytically and one only has numerical results to support
the picture outlined above.

As explained in section \ref{sec:introduction}, there exists the bijection between ensembles of GCDT and a set of graphs with a finite number of faces 
such that the number of local maxima of the distance labeling in GCDT coincides with the number of faces in the set of graphs. 
Coupling matter to GCDT has been studied so far in the sense that the GCDT graphs considered are the graphs with a finite number of faces (including tree-subgraphs), 
e.g. a multicritical one-matrix model for GCDT coupling to hard dimers \cite{Ambjorn:2012zx} and the two-matrix model for coupling to Ising spins \cite{Fuji:2011ce, Sato:2017ccb}.     
The present work is a continuation of the latter, the Ising case.

Apart from the study on coupling of matter to GCDT in the way originally defined, 
it is interesting to work on the matter coupling to the GCDT graphs which are the graphs with a finite number of faces, 
as we have studied in this paper. This is because we may have a chance to observe critical behaviors different from the known:  
In the case of the Ising model coupled to GCDT based on the two-matrix model, 
on the Boulatov-Kazakov critical curve (parametrized by $\theta$) that is characterized by 
infinitely many skeleton vertices and divergent fluctuations of Ising spins, one can recover the critical behavior 
of the $c=1/2$ conformal matter minimally coupled to $2$d gravity 
while going down to the zero critical temperature with tuning $\theta \to 0$, 
one can reach the critical endpoint where the two kinds of critical curves, i.e. the Boulatov-Kazakov critical curve and the critical curve for dominant trees, 
meet, at which we have had a possibility to obtain a new critical behavior. 
In the previous paper \cite{Sato:2017ccb}, it has been shown that the continuum limit around the zero critical temperature can be taken if one scales $\theta$ to be of order $\varepsilon^3$. 
In this paper, we have tried to elucidate the physics around this zero-temperature critical point quantitatively to compute divergent behaviors of the number of vertices and a critical exponent.

We have shown that the continuum limit with the scaling $\theta \sim \varepsilon^3$ proposed by \cite{Sato:2017ccb} 
leads to the fact that the number of skeleton vertices remains finite even in the continuum limit, 
meaning that it is the tree structure that determines the critical behavior, 
i.e. what we have found is the critical behavior identical to that of the Ising model on a $1$d lattice chain or branched polymers, 
which is not satisfactory because this criticality is already known.

One of interesting findings in the present work is the existence of the scaling $\theta \sim \varepsilon^a$ where $0<a<3$ even when coupling to Ising spins.  
With this $a$-dependent scaling, the number of skeleton vertices as well as the number of vertices in the dressed propagators diverges  
as shown by eq.\ \rf{eq:adependentn}; we then have found that one {\it can} obtain a non-trivial scaling 
of the magnetization with the size of the graph, namely eq.\ \rf{ja6}. 
This is indeed a new type of critical behavior in between the Boulatov-Kazakov criticality ($a=0$) and the criticality of the $1$d spin chain ($a=3$).

However, the result is not entirely satisfactory since it essentially reflects the standard Ising spin behavior on sub-lattices consisting of planar
$\phi^3$ skeleton graphs. The main problem seems to be the  character of the ensembles of interpolating graphs. 
They are simply too inhomogeneous, since the Ising spins are non-critical on the tree-subgraphs, except at strictly zero 
temperature, where they are trivially critical, like the $1$d spin chain.

It would be interesting to find better homogeneous ensembles of graphs, i.e. ensembles which have only one well defined fractal 
dimension, and which interpolate between the DT and the GCDT ensembles. 
On these ensembles one might very well find new scaling behavior of the Ising model.

\pagebreak

\section*{Acknowledgement}
We would like to thank  
Timothy Budd, 
Bergfinnur Durhuus, 
Masafumi Fukuma, 
Luca Lionni, 
Tadakatsu Sakai, 
Masaki Shigemori, 
Hidehiko Shimada
and 
Ryo Suzuki 
for fruitful discussions and encouragements.    
YS visited 
the Niels Bohr Institute, Denmark, 
the Radboud University, the Netherlands 
and the Shing-Tung Yau Center of Southeast University, China 
where part of this work was done. He would like to thank all the members there for the kind hospitality. 
JA acknowledges the support from 
the Danish Research Council, via the grant ``Quantum Geometry'', grant no.  7014-00066B.
The work of YS was supported by Building of Consortia for the Development of Human Resources in Science and Technology 
and by JSPS KAKENHI Grant Number 19K14705.





\begin{thebibliography}{40}



\bibitem{Kazakov:1986hu}
  V.~A.~Kazakov,
  ``Ising model on a dynamical planar random lattice: Exact solution,''
  Phys.\ Lett.\ A {\bf 119} (1986) 140.

\bibitem{Boulatov:1986sb}
  D.~V.~Boulatov and V.~A.~Kazakov,
  ``The Ising Model on Random Planar Lattice: The Structure of Phase Transition and the Exact Critical Exponents,''
  Phys.\ Lett.\ B {\bf 186} (1987) 379.
  doi:10.1016/0370-2693(87)90312-1




\bibitem{Ambjorn:1985az}
  J.~Ambj\o rn, B.~Durhuus and J.~Frohlich,
  ``Diseases of Triangulated Random Surface Models, and Possible Cures,''
  Nucl.\ Phys.\ B {\bf 257} (1985) 433.
  doi:10.1016/0550-3213(85)90356-6

\bibitem{Ambjorn:1985dn}
  J.~Ambj\o rn, B.~Durhuus, J.~Frohlich and P.~Orland,
  ``The Appearance of Critical Dimensions in Regulated String Theories,''
  Nucl.\ Phys.\ B {\bf 270} (1986) 457.
  doi:10.1016/0550-3213(86)90563-8


\bibitem{David:1984tx}
  F.~David,
  ``Planar Diagrams, Two-Dimensional Lattice Gravity and Surface Models,''
  Nucl.\ Phys.\ B {\bf 257} (1985) 45.
  doi:10.1016/0550-3213(85)90335-9
  

\bibitem{Billoire:1985ur}
  A.~Billoire and F.~David,
  ``Microcanonical Simulations of Randomly Triangulated Planar Random Surfaces,''
  Phys.\ Lett.\  {\bf 168B} (1986) 279.
  doi:10.1016/0370-2693(86)90979-2


\bibitem{Kazakov:1985ea}
  V.~A.~Kazakov, A.~A.~Migdal and I.~K.~Kostov,
  ``Critical Properties of Randomly Triangulated Planar Random Surfaces,''
  Phys.\ Lett.\  {\bf 157B} (1985) 295.
  doi:10.1016/0370-2693(85)90669-0
 
\bibitem{Boulatov:1986jd}
  D.~V.~Boulatov, V.~A.~Kazakov, I.~K.~Kostov and A.~A.~Migdal,
  ``Analytical and Numerical Study of the Model of Dynamically Triangulated Random Surfaces,''
  Nucl.\ Phys.\ B {\bf 275} (1986) 641.
  doi:10.1016/0550-3213(86)90578-X



\bibitem{Ambjorn:1997di}
  J.~Ambj\o rn, B.~Durhuus and T.~Jonsson,
  ``Quantum Geometry : A Statistical Field Theory Approach,'' 
  Cambridge Monographs on Mathematical Physics, 
  Cambridge University Press, Cambridge, 1997. 



\bibitem{Onsager:1943jn}
 L.~Onsager,
  ``Crystal statistics. 1. A Two-dimensional model with an order disorder transition,''
  Phys.\ Rev.\  {\bf 65} (1944) 117.
  doi:10.1103/PhysRev.65.117


\bibitem{Knizhnik:1988ak}
  V.~G.~Knizhnik, A.~M.~Polyakov and A.~B.~Zamolodchikov,
  ``Fractal Structure of 2D Quantum Gravity,''
  Mod.\ Phys.\ Lett.\ A {\bf 3} (1988) 819.
  doi:10.1142/S0217732388000982



\bibitem{al}
  J.~Ambj\o rn and R.~Loll,
  ``Nonperturbative Lorentzian quantum gravity, causality and topology change,''
  Nucl.\ Phys.\ B {\bf 536} (1998) 407
  doi:10.1016/S0550-3213(98)00692-0
  [hep-th/9805108].

\bibitem{ajl}
  J.~Ambj\o rn, J.~Jurkiewicz and R.~Loll,
  ``A Nonperturbative Lorentzian path integral for gravity,''
  Phys.\ Rev.\ Lett.\  {\bf 85} (2000) 924
  doi:10.1103/PhysRevLett.85.924
  [hep-th/0002050].
  
\bibitem{ajl1}
  J.~Ambj\o rn, J.~Jurkiewicz and R.~Loll,
  ``Dynamically triangulating Lorentzian quantum gravity,''
  Nucl.\ Phys.\ B {\bf 610} (2001) 347
  doi:10.1016/S0550-3213(01)00297-8
  [hep-th/0105267].


\bibitem{Ambjorn:2012jv}
  J.~Ambj\o rn, A.~Goerlich, J.~Jurkiewicz and R.~Loll,
  ``Nonperturbative Quantum Gravity,''
  Phys.\ Rept.\  {\bf 519} (2012) 127
  doi:10.1016/j.physrep.2012.03.007
  [arXiv:1203.3591 [hep-th]].


\bibitem{lollreview}
  R.~Loll,
  ``Quantum Gravity from Causal Dynamical Triangulations: A Review,''
  Class.\ Quant.\ Grav.\  {\bf 37} (2020) no.1,  013002
  doi:10.1088/1361-6382/ab57c7
  [arXiv:1905.08669 [hep-th]].


\bibitem{aal}
  J.~Ambj\o rn, K.~N.~Anagnostopoulos and R.~Loll,
  ``A New perspective on matter coupling in 2-D quantum gravity,''
  Phys.\ Rev.\ D {\bf 60} (1999) 104035
  doi:10.1103/PhysRevD.60.104035
  [hep-th/9904012].
\bibitem{aal1}
  J.~Ambj\o rn, K.~N.~Anagnostopoulos and R.~Loll,
  ``Crossing the c = 1 barrier in 2-D Lorentzian quantum gravity,''
  Phys.\ Rev.\ D {\bf 61} (2000) 044010
  doi:10.1103/PhysRevD.61.044010
  [hep-lat/9909129].
  
  
\bibitem{gaussian}
  J.~Ambj\o rn, A.~T.~Goerlich, J.~Jurkiewicz and H.-G.~Zhang,
  Pseudo-topological transitions in 2D gravity models coupled to massless scalar fields,''
  Nucl.\ Phys.\ B {\bf 863} (2012) 421
  doi:10.1016/j.nuclphysb.2012.05.024
  [arXiv:1201.1590 [gr-qc]].\\
  "A $c =$ 1 phase transition in two-dimensional CDT/Horava?Lifshitz gravity?,''
  Phys.\ Lett.\ B {\bf 743} (2015) 435
  doi:10.1016/j.physletb.2015.03.008
  [arXiv:1412.3873 [gr-qc]].\\
  ``The microscopic structure of 2D CDT coupled to matter,''
  Phys.\ Lett.\ B {\bf 746} (2015) 359
  doi:10.1016/j.physletb.2015.05.026
  [arXiv:1503.01636 [gr-qc]].
  
  

\bibitem{adjspin}
  J.~Ambj\o rn, B.~Durhuus and T.~Jonsson,
  ``A Solvable 2-d gravity model with gamma $>0$,''
  Mod.\ Phys.\ Lett.\ A {\bf 9} (1994) 1221
  doi:10.1142/S0217732394001040
  [hep-th/9401137].
  
 
  
\bibitem{Ambjorn:2007jm}
  J.~Ambj\o rn, R.~Loll, W.~Westra and S.~Zohren,
  ``Putting a cap on causality violations in CDT,''
  JHEP {\bf 0712} (2007) 017
  doi:10.1088/1126-6708/2007/12/017
  [arXiv:0709.2784 [gr-qc]].
  
\bibitem{Ambjorn:2008ta}
  J.~Ambj\o rn, R.~Loll, Y.~Watabiki, W.~Westra and S.~Zohren,
  ``A String Field Theory based on Causal Dynamical Triangulations,''
  JHEP {\bf 0805} (2008) 032
  doi:10.1088/1126-6708/2008/05/032
  [arXiv:0802.0719 [hep-th]].

\bibitem{Ambjorn:2008gk}
  J.~Ambj\o rn, R.~Loll, Y.~Watabiki, W.~Westra and S.~Zohren,
  ``A New continuum limit of matrix models,''
  Phys.\ Lett.\ B {\bf 670} (2008) 224
  doi:10.1016/j.physletb.2008.11.003
  [arXiv:0810.2408 [hep-th]].



\bibitem{Ambjorn:2013csx}
  J.~Ambj\o rn and T.~G.~Budd,
  ``Trees and spatial topology change in CDT,''
  J.\ Phys.\ A: Math.\ Theor.\  {\bf 46} (2013) 315201
  doi:10.1088/1751-8113/46/31/315201
  [arXiv:1302.1763 [hep-th]].
 
  
  
\bibitem{Ambjorn:2008jf}
  J.~Ambj\o rn, R.~Loll, Y.~Watabiki, W.~Westra and S.~Zohren,
  A Matrix Model for 2D Quantum Gravity defined by Causal Dynamical Triangulations,''
  Phys.\ Lett.\ B {\bf 665} (2008) 252
  doi:10.1016/j.physletb.2008.06.026
  [arXiv:0804.0252 [hep-th]].
 
 
   
  

  
  
\bibitem{Ambjorn:2014bga}
  J.~Ambj\o rn, T.~Budd and Y.~Watabiki,
  ``Scale-dependent Hausdorff dimensions in 2d gravity,''
  Phys.\ Lett.\ B {\bf 736} (2014) 339
  doi:10.1016/j.physletb.2014.07.047
  [arXiv:1406.6251 [hep-th]].


\bibitem{Fuji:2011ce}
  H.~Fuji, Y.~Sato and Y.~Watabiki,
  ``Causal Dynamical Triangulation with Extended Interactions in 1+1 Dimensions,''
  Phys.\ Lett.\ B {\bf 704} (2011) 582
  doi:10.1016/j.physletb.2011.09.032
  [arXiv:1108.0552 [hep-th]].
  
    



\bibitem{Sato:2017ccb}
  Y.~Sato and T.~Tanaka,
  ``Criticality at absolute zero from Ising model on two-dimensional dynamical triangulations,''
  Phys.\ Rev.\ D {\bf 98} (2018) no.2,  026026
  doi:10.1103/PhysRevD.98.026026
  [arXiv:1710.03402 [hep-th]].



 
 
 
  
\bibitem{Ambjorn:1992rp}
  J.~Ambj\o rn, B.~Durhuus, T.~Jonsson and G.~Thorleifsson,
  ``Matter fields with $c > 1$ coupled to $2$-d gravity,''
  Nucl.\ Phys.\ B {\bf 398} (1993) 568
  doi:10.1016/0550-3213(93)90604-N
  [hep-th/9208030].






\bibitem{hausdorff}
  H.~Kawai, N.~Kawamoto, T.~Mogami and Y.~Watabiki,
  "Transfer matrix formalism for two-dimensional quantum gravity and fractal structures of space-time,''
  Phys.\ Lett.\ B {\bf 306} (1993) 19
  doi:10.1016/0370-2693(93)91131-6
  [hep-th/9302133].\\
  J.~Ambj\o rn and Y.~Watabiki,
  ``Scaling in quantum gravity,''
  Nucl.\ Phys.\ B {\bf 445} (1995) 129
  doi:10.1016/0550-3213(95)00154-K
  [hep-th/9501049].\\
  J.~Ambj\o rn, J.~Jurkiewicz and Y.~Watabiki,
  ``On the fractal structure of two-dimensional quantum gravity,''
  Nucl.\ Phys.\ B {\bf 454} (1995) 313
  doi:10.1016/0550-3213(95)00468-8
  [hep-lat/9507014].


\bibitem{Ambjorn:2012zx}
J.~Ambjorn, L.~Glaser, A.~Gorlich and Y.~Sato,
``New multicritical matrix models and multicritical 2d CDT,''
Phys. Lett. B \textbf{712} (2012), 109-114
doi:10.1016/j.physletb.2012.04.047
[arXiv:1202.4435 [hep-th]].



\end{thebibliography}
\end{document}